\documentclass[epj]{svjour}
\usepackage{graphics}
\begin{document}
\title{Electroexcitation of the Roper resonance from CLAS data}
\author{Inna Aznauryan\inst{1,2} \and Volker Burkert\inst{1}}                    
\institute{Thomas Jefferson National Accelerator Facility, 
Newport News, Virginia 23606, USA\and 
Yerevan Physics Institute, 375036 Yerevan, Armenia}
\date{October 27, 2007}
\abstract
{The helicity amplitudes of the electroexcitation
of the Roper resonance on proton are extracted
at $1.7<Q^2<4.2~GeV^2$ from recent high precision
JLab-CLAS  cross sections data and longitudinally polarized
beam asymmetry for $\pi^+$ electroproduction on protons.
The analysis is made using two approaches, dispersion relations
and unitary isobar model, which give consistent results.
It is found that the transverse helicity amplitude
for the $\gamma^* p\rightarrow P_{11}(1440)$ transition,
which is large and negative at $Q^2=0$,
becomes large and positive at $Q^2\simeq 2~GeV^2$,
and then drops slowly with $Q^2$. Longitudinal helicity amplitude, that was
previously found from CLAS data as large and positive at
$Q^2=0.4,~0.65~GeV^2$, drops
with  $Q^2$. These results
rule out the presentation of $P_{11}(1440)$ as a $q^3G$
hybrid state, and provide strong evidence in favor
of this resonance as a first radial excitation
of the $3q$ ground state.
} 
\maketitle
In this talk we report our results on the electroexcitation
of the Roper resonance ($P_{11}(1440)$) extracted from a large body
of CLAS data on
differential cross sections and polarized beam asymmetries
for the process $ep\rightarrow en\pi^+$
in the range of invariant hadronic mass
$W=1.15-1.69~GeV$ and photon virtuality
$Q^2=1.7-4.2~GeV^2$ with full azimuthal and polar
angle coverage \cite{Park}. Combined with the information
obtained from the previous CLAS
data at $Q^2=0.4,~0.65~GeV^2$ \cite{Azn04,Azn065} and that at
$Q^2=0$ \cite{PDG}, these results give us knowledge of the Roper
electroexcitation in wide $Q^2$ range and 
allow us to draw quite definite conclusions on the nature
of $P_{11}(1440)$.

It is known that the structure of the Roper resonance
has attracted special attention since its discovery,
because the simplest and most natural assumption that this is a
first radial excitation of the $3q$ ground state led to
the difficulties in the description of the resonance.
To deal with shortcomings of the quark model,
alternative, as well extended descriptions of $P_{11}(1440)$ were
developed:
as a hybrid ($q^3G$) state  \cite{Li1,Li2},
that of a quark core dressed by a meson cloud
\cite{Cano1,Cano2}, a dynamically generated $\pi N$ resonance \cite{Krewald},
and presentations that include $3q-q\bar{q}$ components,
in particular, strong $\sigma N$ component (see Ref. \cite{Dillig}
and references therein).

The $Q^2$ dependence of the electromagnetic transition
form factors is highly sensitive to different
descriptions of the Roper state.
However, until recently, the data base needed to measure
these form factors at relatively high $Q^2$
was almost exclusively based on $\pi^0$
production, and was very limited in kinematical coverage.
Also, the $\pi^0 p$ final state is dominated by the nearby isospin
3/2 $\Delta(1232)$ resonance, whereas the isospin 1/2 Roper state
couples more strongly to the $\pi^+ n$ channel.
The data \cite{Park} allow us to 
extract the $\gamma^* p\rightarrow P_{11}(1440)$ 
helicity amplitudes at $Q^2=1.7-4.2~GeV^2$, and therefore
to obtain complete picture
of the electroexcitation of the Roper resonance in 
a wide $Q^2$ region. 

The approaches we use are fixed-$t$
dispersion relations (DR) and unitary isobar model (UIM),
which both
were successfully employed in Refs. \cite {Azn04,Azn065,Azn0} to analyze
photoproduction and low $Q^2$ electroproduction of pions.
These approaches were presented and discussed
in Refs. \cite{Azn0}.                                                                                                           

The imaginary parts of the amplitudes in both approaches
are determined mainly by $s$- channel resonance contributions
which we parameterize in the Breit-Wigner form
with energy-dependent widths \cite{Azn0,Drechsel}. The exception
was made  for
the $P_{33}(1232)$ resonance which was treated
in a special way. According to the phase-shift analyses 
of the $\pi N$ scattering,
the $\pi N$ amplitude corresponding to 
the $P_{33}(1232)$ resonance is elastic
up to $W=1.43~GeV$ (see, for example, the
latest GWU(VPI) analyses \cite{GWU1,GWU2}).
In combination with DR and Watson's theorem, this provides 
strict constraints on the multipole amplitudes
$M_{1+}^{3/2}$, $E_{1+}^{3/2}$, $S_{1+}^{3/2}$
that correspond to the $P_{33}(1232)$ resonance.
In particular, as it was shown in Ref. \cite{Azn0},
the shape of $M_{1+}^{3/2}$
is close to that at $Q^2=0$
from the GWU(VPI) analysis \cite{GWU3}.
This constraint on the large 
$M_{1+}^{3/2}$ amplitude plays an important role in the reliable
extraction of the $P_{11}(1440)$ electroexcitation
amplitudes, because the $P_{33}(1232)$ and $P_{11}(1440)$ states
are strongly overlapping.

We have taken into account
all  resonances from the first, second, and third resonance
regions. These are 4-and 3-star resonances
$P_{33}(1232)$, $P_{11}(1440)$,
$D_{13}(1520)$, $S_{11}(1535)$,
$P_{33}(1600),~$ $S_{31}(1620),~$
$S_{11}(1650),~$
$D_{15}(1675),~~~~~$ 
$F_{15}(1680)$,
$D_{13}(1700)$, $D_{33}(1700)$,
$P_{11}(1710)$, and
$P_{13}(1720)$.
The masses, widths, and  $\pi N$ branching
ratios  of these resonances were taken equal
to the mean values of the data presented in the Review of Particle
Physics (RPP) \cite{PDG}. 
At each $Q^2$, we have made two kinds of fits
in both approaches:
(i) The magnitudes of the helicity amplitudes
corresponding to all resonances
listened above were fitted.
(ii) The transverse amplitudes for
the members of the multiplet $[70,1^-]$: $S_{31}(1620)$,
$S_{11}(1650)$,
$D_{15}(1675)$, $D_{13}(1700)$, and $D_{33}(1700)$,
were fixed according to
the single quark transition model  \cite{SQTM},
which relates these amplitudes to those for
$D_{13}(1520)$ and $S_{11}(1535)$;
the longitudinal amplitudes of these resonances
and the amplitudes of the resonances
$P_{33}(1600)$,
$P_{11}(1710)$, and
$P_{13}(1720)$ were taken equal to 0. 
It turned out that the results obtained
for $P_{33}(1232)$, $P_{11}(1440)$,
$D_{13}(1520)$, and $S_{11}(1535)$ in the two fits
are close to each other. The amplitudes
of the Roper resonance
presented below are the average values of the results
obtained in these fits.

In Fig. 1, we present the comparison of our results
with the experimental data
for the Legendre moments  of
the structure function
$\sigma_T+\epsilon \sigma_L$
at $Q^2=2.05~GeV^2$ \cite{Park}.
The  Legendre moment $D_0^{T+\epsilon L}$ does not
contain interference of different multipole amplitudes and
is related
to the sum of  squares of these amplitudes.
The resonance behavior of the multipole amplitudes is revealed
in $D_0^{T+\epsilon L}$
in the form of enhancements. Resonance structures
related to the narrow resonances
$P_{33}(1232)$, $D_{13}(1520)$, and $S_{11}(1535)$
are clearly  seen in
$D_0^{T+\epsilon L}$.
There is a shoulder between the $\Delta$
and $1.5~GeV$ peaks, which is related to the broad Roper resonance.
To demonstrate this we present in Fig. 1
by dotted curves the results obtained by
switching off the $P_{11}(1440)$ resonance
from the DR results.
To stress the advantage of the investigation of
the Roper resonance in the reaction $\gamma^* p \rightarrow \pi^+ n$,
we note that for this reaction,
the relative contribution
of $P_{11}(1440)$ in comparison with $P_{33}(1232)$
in $D_0^{T+\epsilon L}$
is 4 times larger than for $\gamma^* p \rightarrow \pi^0 p$.                                                                                               
\begin{figure}
\resizebox{0.48\textwidth}{!}{%
  \includegraphics{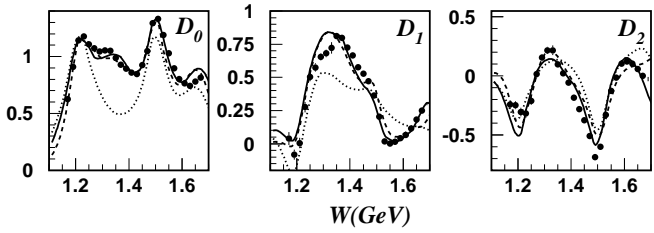}
}
\caption{Experimental data for the Legendre moments
of the structure function $\sigma_T+\epsilon \sigma_L$ at
$Q^2=2.05~GeV^2$ \cite{Park} in comparison with our
results; the units are $\mu b/sr$. Solid and dashed curves correspond to the
DR and UIM results, respectively.
Dotted curves are obtained by
switching off the $P_{11}(1440)$ resonance
from the DR results.
}
\label{fig:1}       
\end{figure}

\begin{table}[t]
\begin{tabular}{|c|c|c|}
\hline
$Q^2~(GeV^2)$&$A_{1/2}~~~~~~~~~~~~~~S_{1/2}$\\
\hline
&DR\\
\hline
1.72&$72.5\pm 1.0\pm 4.3~~~~24.8\pm 1.4\pm 5.3$\\
\hline
2.05&$72.0\pm 0.9\pm 4.2~~~~21.0\pm 1.7\pm 5.0$\\
\hline
2.44&$50.0\pm 1.0\pm 3.2~~~~9.3\pm 1.3\pm 4.1$\\
\hline
2.91&$37.5\pm 1.1\pm 2.8~~~~9.8\pm 2.0\pm 2.3$\\
\hline
3.48&$29.6\pm 0.8\pm 2.7~~~~4.2\pm 2.5\pm 2.3$\\
\hline
4.16&$19.3\pm 2.0\pm 3.9~~~~10.8\pm 2.8\pm 4.5$\\
\hline
&UIM\\
\hline
1.72&$58.5\pm 1.1\pm 4.2~~~~26.9\pm 1.3\pm 5.3$\\
\hline
2.05&$62.9\pm 0.9\pm 3.3~~~~15.5\pm 1.5\pm 4.9$\\
\hline
2.44&$56.2\pm 0.9\pm 3.2~~~~11.8\pm 1.4\pm 4.1$\\
\hline
2.91&$42.5\pm 1.1\pm 2.8~~~~13.8\pm 2.1\pm 2.3$\\
\hline
3.48&$32.6\pm 0.9\pm 2.6~~~~14.1\pm 2.4\pm 2.0$\\
\hline
4.16&$23.1\pm 2.2\pm 4.8~~~~17.5\pm 2.6\pm 5.5$\\
\hline
\end{tabular}
\caption{\label{tab1}The
$\gamma^* p \rightarrow P_{11}(1440)$
helicity amplitudes
(in $10^{-3}GeV^{-1/2}$ units) found from the analysis
of  $\pi^+$ electroproduction
data \cite{Park} using DR and UIM.
First uncertainty has statistical nature, it was obtained
in the fitting procedure. Second uncertainty is systematic one;
it is connected with the averaging procedure
of the results obtained in two kinds of fits,
discussed in the text, and with the uncertainties of the background
caused by the uncertainties of the nucleon, pion and
$\rho(\omega)\rightarrow\pi\gamma$ form factors.
}
\end{table}

\begin{figure}
\resizebox{0.48\textwidth}{!}{
  \includegraphics{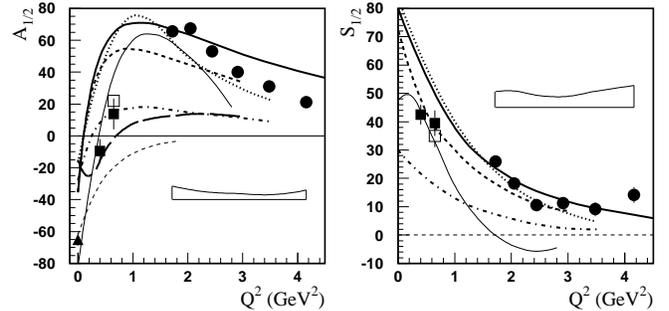}
}
\caption{Helicity amplitudes
for the  $\gamma^* p \rightarrow P_{11}(1440)$ transition
(in $10^{-3}GeV^{-1/2}$ units).
Full circles are the average values of
our results obtained from
the analysis
of  $\pi^+$ electroproduction
data \cite{Park} using DR and UIM.
The bands present the systematic uncertainties wich are caused
by the averaging procedure and by systematic uncertainties
mentioned in the caption to Table 1.
Full boxes are  the results obtained
from CLAS data \cite{Azn04,Joo1,Joo2,Joo3,Egiyan};
open boxes present the results of the combined
analysis of CLAS
single $\pi$ and $2\pi$ electroproduction
data \cite{Azn065}.
Full triangle at 
$Q^2=0$ is the RPP estimate \cite{PDG}.
Thick curves correspond to
the light-front relativistic quark models:  dotted, dashed,
dash-dotted, long-dashed, and solid curves are
from Refs. \cite{Weber,Capstick,Simula,Riska,Quark},  respectively.
Thin solid curves are the predictions obtained for the
Roper resonance treated as
a quark core dressed by a meson cloud \cite{Cano1,Cano2}.
Thin dashed curves are
obtained assuming that $P_{11}(1440)$
is a $q^3 G$ hybrid state \cite{Li2}.
}
\label{fig:2}       
\end{figure}
We now discuss the results for the
$\gamma^* p \rightarrow P_{11}(1440)$
helicity amplitudes presented 
in Table \ref{tab1} and Fig. 2.

It can be seen that the results
obtained using DR and UIM
are close to each other.
As the non-resonant backgrounds of these approaches are
built in conceptually different ways,
we conclude that the model uncertainties
of the obtained results are small.
                                                                                                          
Combined with the information obtained from
the previous CLAS data at $Q^2=0.4,~0.65~GeV^2$
\cite{Azn04,Azn065,Joo1,Joo2,Joo3,Egiyan},
and that at $Q^2=0$ \cite{PDG}, new results show
nontrivial behavior
of the transverse helicity
amplitude $A_{1/2}$:
being large and negative at $Q^2=0$,
it crosses zero between $Q^2=0.4$ and $0.65~GeV^2$
and becomes large and positive
at $Q^2\simeq 2~GeV^2$.
Further with
increasing $Q^2$, this amplitude drops smoothly in magnitude.
The longitudinal helicity amplitude $S_{1/2}$, which is large and
positive at small $Q^2$, drops smoothly with increasing $Q^2$.

In Fig. 2, we compare our results
with model predictions.
These are (i) quark model predictions 
\cite{Weber,Capstick,Simula,Riska,Quark}
where the $P_{11}(1440)$ is described
as the first radial excitation of the $3q$ ground state;
(ii) those assuming the
$P_{11}(1440)$ is
a hybrid state  \cite{Li2}; and (iii)
the results for the
Roper resonance treated as
a quark core (which is a radial excitation of the $3q$ ground state)
dressed by a meson cloud \cite{Cano1,Cano2}.
                                                                                                          
It is known that with increasing $Q^2$,
when the momentum transfer becomes larger
than the masses of the constituent quarks, a relativistic
treatment of the electroexcitation of the nucleon resonances,
which is important already at $Q^2=0$,
becomes crucial.
The consistent way to realize the relativistic
treatment of the
$\gamma^* N \rightarrow N^*$  transitions
is to consider
them in the LF dynamics.
In Fig. 2 we compare our results with the predictions
of the LF quark models \cite{Weber,Capstick,Simula,Riska,Quark}.
                                                                                                          
All LF approaches \cite{Weber,Capstick,Simula,Riska,Quark}
give good description of the nucleon form factors, however,
the predictions for the $\gamma^* N \rightarrow P_{11}(1440)$
helicity amplitudes are quite different.
This is caused by the large sensitivity
of these amplitudes to
the $N$ and $P_{11}(1440)$ wave functions \cite{Quark}.
                                                                                                          
The approaches \cite{Weber,Capstick,Simula,Riska,Quark}
fail to describe the value of the
transverse amplitude $A_{1/2}$ at $Q^2=0$.
This can be an indication of a large meson cloud
contribution to the $\gamma^* p \rightarrow P_{11}(1440)$
which is expected to be significant at small $Q^2$.
As a confirmation of this assumption one can consider
the results of Refs. \cite{Cano1,Cano2} where
this contribution
is taken into account,
and a good description of the helicity amplitudes
is obtained at small $Q^2$.
                                                                                                          
In spite of
differences, all LF predictions
for the $\gamma^* p \rightarrow P_{11}(1440)$
helicity amplitudes
have common features
which agree
with the results extracted from the experimental data:
(i) the sign of the transverse amplitude
$A_{1/2}$ at $Q^2=0$
is negative,
(ii) the sign
of the longitudinal
amplitude $S_{1/2}$ is positive,
(iii) all LF approaches predict the sign change 
of the transverse amplitude $A_{1/2}$ at small $Q^2$.
We take this qualitative agreement as the evidence in the favor
of the $P_{11}(1440)$ resonance as a radial excitation
of the $3q$ ground state. Final confirmation of this conclusion requires
complete simultaneous description of the nucleon
form factors and the  $\gamma^* p \rightarrow P_{11}(1440)$ amplitudes.
This will allow us to  find the magnitude of the meson cloud contribution, and
to better specify the $N$ and $P_{11}(1440)$ wave functions.
                                                                                                          
The results of Refs. \cite{Li1,Li2}, where
$P_{11}(1440)$ is treated as
a hybrid state,  are obtained via non relativistic calculations.
Nevertheless the suppression
of the longitudinal amplitude $S_{1/2}$ has its physical
origin
in the fact that the longitudinal
transition operator for the vertex $\gamma q\rightarrow q G$
requires both spin and angular momentum flip by one unit,
while the angular momenta of quarks in the $N$ and $P_{11}(1440)
\equiv q^3 G$ are equal to 0.
This makes this result practically independent
of relativistic effects.
The suppression of the longitudinal
amplitude $S_{1/2}$ 
strongly disagrees with the experimental results.
                                                                                                          
In summary, for the first time the transverse and longitudinal helicity
amplitudes of the  $\gamma^* p \rightarrow P_{11}(1440)$ transition
are extracted from experimental data
at high $Q^2$. The results are obtained from differential
cross sections and longitudinally polarized
beam asymmetry for $\pi^+$ electroproduction on protons
at $W=1.15-1.69~GeV$ \cite{Park}.
The data were analyzed using two
conceptually different approaches, DR and UIM,
which give consistent results.
                                                                                                          
The strong longitudinal coupling rules out the presentation of
the Roper resonance as a $q^3G$ hybrid state, while
comparison with quark model predictions
provides strong evidence in favor
of $P_{11}(1440)$ as a first radial excitation
of the $3q$ ground state.

$\bf{Acknowledgements}$. The work was supported
by U.S. DOE contract DOE/0R/23177-0182.

\end{document}